# Heterogeneously Integrated ITO Plasmonic Mach-Zehnder Interferometric Modulator on SOI

Rubab Amin[1], Rishi Maiti[1], Yaliang Gui[1], Can Suer[1], Mario Miscuglio[1], Elham Heidari[2], Jacob B. Khurgin[3], Ray T. Chen[2], Hamed Dalir[4], and Volker J Sorger[1,*]

[1]Department of Electrical and Computer Engineering, George Washington University, Washington, DC 20052, USA
[2]Microelectronics Research Center, Electrical and Computer Engineering Department, University of Texas at Austin, Austin, Texas 78758, USA
[3]Department of Electrical and Computer Engineering, Johns Hopkins University, Baltimore, Maryland 21218, USA
[4]Optelligence LLC, Alexandria, Virginia, 22302, USA
*Corresponding author: sorger@gwu.edu

## Abstract

**Densely integrated active photonics is key for next generation on-chip networks for addressing both footprint and energy budget concerns. However, the weak light-matter interaction in traditional active Silicon optoelectronics mandates rather sizable device lengths. The ideal active material choice should avail high index modulation while being easily integrated into Silicon photonics platforms. Indium tin oxide (ITO) offers such functionalities and has shown promising modulation capacity recently. Interestingly, the nanometer-thin unity-strong index modulation of ITO synergistically combines the high group-index in hybrid plasmonic with nanoscale optical modes. Following this design paradigm, here, we demonstrate a spectrally broadband, GHz-fast Mach-Zehnder interferometric modulator, exhibiting a high efficiency signified by a miniscule $V_\pi L$ of 95 V·µm, deploying an one-micrometer compact electrostatically tunable plasmonic phase-shifter, based on heterogeneously integrated ITO thin films into silicon photonics. Furthermore we show, that this device paradigm enables spectrally broadband operation across the entire telecommunication near infrared C-band. Such sub-wavelength short efficient and fast modulators monolithically integrated into Silicon platform open up new possibilities for high-density photonic circuitry, which is critical for high interconnect density of photonic neural networks or applications in GHz-fast optical phased-arrays, for example.**

## Introduction

Indium tin oxide (ITO), belonging to the class of transparent conductive oxides, is a material extensively adopted in high-tech industry such as in touchscreen displays of smartphones or contacts for solar cells. Recently, ITO has been explored for electro-optic (EO) modulation using its free-carrier dispersive effect enabling unity-strong index modulation [1-4]. The advantages of using ITO as opposed to traditional materials (e.g. Si) in photonic integrated circuit (PIC) modulation is manifold: a) the carrier concentration in ITO can be 3-4 orders of magnitude higher compared to that of Si allowing more carriers to tune the dispersion behavior, this coupled with the capacitive gating schemes for altering ITO carrier concentrations as opposed to p-n junction based Si schemes, in turn, means high capacitance enabled ITO modulators can obtain higher charge modulation than Si for the same voltage; b) the effect of the tuning carrier concentration on the optical parameters, i.e. index alteration, per unit applied bias, is in fact more dramatic compared to Si due to the lower permittivity of ITO compared to Si ($\partial n = \partial \epsilon^{1/2} \sim \partial \epsilon / 2\epsilon^{1/2}$); c) the presence of an epsilon near zero (ENZ) region in the tolerable carrier concentration range within electro-static gating constraints can enhance attainable modulation effects; d) the potential complementary-metal-oxide-semiconductor (CMOS) compatibility, i.e. the foundry adaptability of ITO allow for a seamless integration with the mature Si process.

Thus, when taken together, ITO stands out from the crowd of next-generation active modulation materials. The latter is the "actual" reason for recent interest in usage of ITO in modulation. However, GHz-fast modulation capability using ITO is forthcoming in recent literature – a feature we show herein. Given the ubiquitous usage of phase-shifter technologies, such as in data communication, optical phased arrays, analog and RF photonics, sensing etc.; here we focus on a Mach-Zehnder interferometer (MZI) based modulator to demonstrate a comprehensive platform of heterogeneous integration of ITO-based opto-electronics into silicon PICs. Since for phase-shifters only the real-part of the optical refractive index ($n$) is of interest, in previous studies we have shown the interplay between a selected optical mode (e.g. photonic bulk vs. plasmonic) and the material's figure of merit ($\Delta n/\Delta \alpha$), where $\alpha$ is the optical loss, is a direct result from Kramers-Kronig (K-K) relations [5]. Additionally, ITO can be selectively prepared (via process conditions [6]) for operating in either an $n$-dominant or $\alpha$-dominated region [5]. Using this approach, we recently showed a photonic-mode ITO-oxide-Si MZI on silicon photonics characterized by a $V_\pi L$ = 0.52 V·mm [3], and a plasmonic version deploying a lateral gate exhibiting a $V_\pi L$ = 0.063 V·mm [7]. Indeed, a plasmonic mode enables a strong light-matter-interaction (e.g. extrinsic slow-light effect), which, when superimposed with ITO's intrinsic slow-light effect, proximal epsilon-near-zero (ENZ) effects allowing higher light confinement with tuning in the modal structure [8], enables realization of just 1-5 μm short phase-shifters [5]. The device-advantage of such micrometer-compact opto-electronics are small capacitances, in the order of ~fF, enabling low power consumption and small RC delays for fast signal reconfiguration, as shown here. Note, ENZ operation incurs high losses detrimental for phase shifters; as such, the desired operation region is here adequately close-to, but not at the high-loss ENZ (ENZ located in $\alpha$-dominant region) [5].

One can certainly assert that there are applications where established EO materials such as lithium niobate (LiNbO$_3$, LN), especially bonded to Si, has particular modulation advantages but applications requiring short haul, integrated, CMOS compatible and on-chip functionalities can synergistically benefit from the adoption of heterogeneous monolithic integration of ITO. It is also worthwhile to mention here that, the demonstrated capabilities of ITO as a strong nonlinear material can designate desirability as one can combine many different functionalities on the same chip with the same active material [9,10]. In fact, unlike LN optoelectronics requiring careful crystal orientation control [11], ITO thin films are crystal-orientation independent and feature intrinsically homogeneous uniform optical characteristics. The weak EO effect of LN leading to cm-long devices does not only obligate macroscopic large dimension, but more importantly requires delicate driver design (impedance matching in traveling wave circuitry), and, when combined, are prohibitive for dense chip integration which is one key disadvantage of silicon-based PIC platforms. The lurking concern with LN modulators is geometry – the RF field being not tightly confined is wasteful, both electrodes placed on opposite sides of waveguide with the active LN material necessarily widens the gap requiring large voltage for the E-field manipulations; whereas for the ITO case one of the electrodes is the active layer itself tightening the gap for reduced voltage manipulations. ITO, in conjunction with a plasmonic optical metal-oxide-semiconductor (MOS) mode, on the other hand, can harness the full opportunities of electrostatic gating that the semiconductor industry has perfected over the last ½ century. In LN-optoelectronics, however, a metal contact cannot be placed in close proximity to the waveguide, which is required for efficient electrostatics, because the so-incurred loss-per-unit-length would be detrimental. Furthermore, ITO thin films are crystal-orientation independent, while LN thin films require careful crystal orientation control (x-cut, z-cut, etc.); thus, small variation in the Pockels coefficient in LN-modulators can cause significant alterations in device performance (for example, in a ring geometry or bent waveguides). In contrast, ITO features intrinsically homogeneous uniform optical characteristics leading to high orders of process control. Also, the selective etching process in microstructuring LN films is quite delicate and arduous; whereas heterogeneous integration of ITO thin films on Si is rather straightforward. Microstructured LN architectures are not compatible with the mature Si process flow, whilst the semiconductor industry already process ITO in their foundries for a plethora of applications including display technology and transparent contacts. Indeed ITO is already processed routinely by the smart-phone display industry (Apple, Samsung) and solar-cell industry. When compared, LN seems the odd-one-out, and ITO bears much potential for market adoption specifically in short haul on-chip applications such as neuromorphic photonics [12].

Here, we introduce and demonstrate an ITO-plasmon-based phase-shifter heterogeneously integrated into a silicon photonic MZI delivering GHz-fast modulation while being spectrally broadband, thus opening up opportunities for multi-spectral operation. Refraining to exploit feedback from a resonator has several advantages: a) no spectral alignment to the pump laser is required; b) Thus, no heaters are needed for thermal tuning, which would raise the power consumption per-device into the ~nJ/bit levels [13]; c) Heat spreads ~100 μm's across the PIC, thus packaging density can be significantly improved; d) The photon-lifetime of high-quality ($Q$) cavities can limit RF modulation speeds due to the long photon lifetime, which is not the case here. Our ITO hybrid plasmonic MZI modulator can exploit efficient electrostatics to its fullest potential leading to CMOS-compatible (~1-2 V) small drive voltages when optimized in contrast to LN-based schemes. Note, despite the high losses per unit length the total loss of our approach is actually comparable to silicon photonics devices, because the device is just mere 1-3 μm short while enabling a $V_\pi L$ being 2-3 orders of magnitude smaller.

## Figure of Merit (FOM) for MZIs

Phase modulators invariably require some method of interferometric scheme such as cavity feedback or MZI-based arrangements to compare the actively modulated optical phase change with a reference beam in order to induce a change in optical transmitted amplitude via interference. As such, these schemes intrinsically suffer from an extended footprint conundrum, as opposed to electro absorption schemes that can be implemented rather straightforwardly in a linear waveguide arrangement. In MZIs, the product of the half-wave voltage and the active modulator length, $V_\pi L$, is a key figure of merit (FOM), since both elemental parameters exhibit a tradeoff between obtaining π-phase shifts at the optical output with increased device length or applied voltage. As such, continuous efforts to minimize $V_\pi L$ in MZI schemes have succeeded in realizing miniscule FOMs (10's of V·μm) over the years (Table 1).

Lithium niobate-based MZIs are commercially available featuring rather high FOMs ($V_\pi L = 10^4$-$10^5$ V·μm) due to their field-based modulation mechanism characterized by the weak Pockels effect necessitating extremely long device lengths in the range of 1-10 centimeters. Improvements in performance are obtained with the QCSE in III-V [31,33,35] or enhanced Pockels effect in emerging materials such as polymers [28,32,37-39]. Augmented light-confinement for improving optical and RF mode overlap (Γ) with the active material in plasmonic structures have also been sought [37-39] (Table 1). This work focuses on an experimentally demonstrated plasmonic ITO based MZI device and the achieved FOM in our results suggests that our performance is on-par with recent exemplary works despite being the first of its kind.

The devices exhibiting a high-performance FOM (i.e. reduced $V_\pi L$) amount to plasmonics, integration of organic/polymer materials, III-V quantum well structures, etc (Table 1). While many of these schemes essentially offer acceptable performance, they are mostly difficult to integrate in the mature Si process. Our choice of ITO as the active modulation material can avail ease of fabrication and potential CMOS integration [40].

## Device Design

A symmetrical passive 1×1 MZI structure on a Silicon on insulator (SOI) platform is chosen as the underlying structure to distinguish the interference pattern at the output confering modulation effects from the active device only. A symmetrical MZI structure in this context refers to the same length for both Mach-Zehnder arms and commensurate 50:50 Y-junctions on both sides. This essential design choice refrains the optical output from running astray as different ratio in the Y-junctions can introduce chirp effects in modulation. Fabrication imperfections such as sidewall roughness, surface defects and alignment issues inherently deter from achieving perfectly symmetrical MZI structures. Nevertheless, this design choice is enforced to minimize such effects and curtail the optical path length difference in both the arms of the MZI, to distinguish phase variability only conferring active modulation effects.

Subsequent process steps towards the active device include depositing a 10 nm thin film of ITO on the device region using ion beam deposition (IBD) and a top plasmonic metal (Ti/Au) stack on top of a portion of one arm separated by an oxide layer to facilitate gating (Fig. 1). Since the modulation efficiency (ER/$V_{pp}$) is improved for enhanced electrostatics, we use a relatively high-k dielectric, $Al_2O_3$, for the gate oxide of 20 nm using atomic layer deposition (ALD). Note, IBD allows synergies for processing ITO as it can yield pinhole-free dense crystalline films with high uniformity, and allows for a room temperature process not annealing the ITO as a byproduct of the process (i.e. no activation of Sn carriers to facilitate electrostatic EO modulation as opposed to applications in contacts or touch screen displays). Incidentally, IBD technologies are advantageous for nanophotonic device fabrication due to their precise controllability of material properties such as microstructure, non-stoichiometry, morphology, crystallinity, etc. [41,42].

Nevertheless, highly-index tunable materials alike ITO are inherently accompanied by increased absorption as a byproduct of modulation dictated by the K-K relations. This absorption in free-carrier dispersion based materials (e.g. Si, ITO) directly arises from the well-known K-K relations and poses a tradeoff for utilizing such highly tunable active modulation materials [43-45]. As such, the active ITO in one arm of the MZI imposes a modulating absorption (loss) component, whereas the other (un-modulated) arm outputs a higher optical power as a result. This, in turn, produces an arm-loss imbalance in the MZI, thus limiting the achievable modulation dynamic range as differential loss between the arms due to carrier depletion can limit the extinction ratio (ER) [46]. It originates from the complex part of the optical phase and leads to degraded optical signal fidelity (i.e. phase errors at the output and alter frequency chirps) [47].

Intrinsically, any induced change in the complex refractive index (real or imaginary part) of the active material impacts the counterpart as a byproduct (i.e. K-K relations), thus the loss imbalance can alter during MZM operation between the ON/OFF states. Improvements in the arm loss imbalance can be obtained by tuning the arm losses statically. A challenge in using an inherently lossy material such as ITO, in an interferometric scheme similar to the MZ configuration to achieve satisfactory modulation depth, is to match the amplitude of the optical signal (i.e. loss) in both arms of the MZ. Essentially, the interference from both arms at the output terminal should converge, referred to hereafter as balancing. As such, adjusting for both ON and OFF states of modulator operation, we calculate the desired length of the passive metal contact, $L_b$ on the un-modulated arm of the MZM (Fig. 1) as it imposes necessary loss on the un-modulated arm to facilitate modulation depth, i.e. balancing the field loss imbalance in both the arms [3].

## Material and Modal Properties

The underlying modulation mechanism for ITO is free-carrier dispersion from an applied bias. The Drude model characterizes the ITO material response well in the near infra-red region ($\lambda > 1$ µm). The relative permittivity of free electron plasma in the Drude approximation can be written as $\epsilon_r = \epsilon_\infty - [N_c q^2 / \epsilon_0 m^* (\omega^2 + i\gamma\omega)]$; where $\epsilon_\infty$ is the dielectric constant of the undoped semiconductor also referred to as the high-frequency or "background" permittivity, $q$ is the electronic charge, $m^*$ is the conduction effective mass, $\omega$ is the angular frequency, and $\gamma = 1/\tau$ is the collision frequency or scattering rate of the free carriers quantified by the detuning from any interband resonances thereof. Our deposited ITO thin films exhibit similar optical indices (real and imaginary parts) as obtained from Drude model fits in previous works [43-45]. Spectroscopic ellipsometry of deposited thin films show a decreasing real part, $n$ and an increasing imaginary part, $\kappa$ of the material index near our operating wavelength, $\lambda = 1550$ nm (Fig. 2a), which is expected from the K-K characteristics [5,44].

Parameters obtained from variable angle spectroscopic ellipsometry ($\Psi$ and $\Delta$) and subsequent fitting derived analysis revealed relevant electrical and optical properties of our ITO thin film as deposited (Fig. 2). Ellipsometric measurements address the change in the polarization ($p$- and $s$-polarizations) of the reflected light from our deposited ITO thin film through a ratio of the complex Fresnel coefficients ($\rho = \tilde{r}_p / \tilde{r}_s = \tan(\Psi) \exp(i\Delta)$) where $\Psi$ relates to an amplitude ratio and $\Delta$ relates to the phase variation from the incident phase [48]. A general oscillator-based approach in fitting assures relevant physically realizable (K-K consistent) parameters extraction (see methods). A Drude fit in our wavelengths of interest (i.e. NIR region) indicates the as-deposited carrier concentration, $N_c$ of 3.13 × $10^{20}$ cm$^{-3}$. Corresponding parameters such as scattering time, $\tau$ of 4.72±0.01 fs; resistivity, $\rho$ of 8.41×$10^{-4}$ Ω·cm and mobility, $\mu$ of 23.72±0.06 cm$^2$/V-s are also extracted from the ellipsometric data and reinforced with further metrology such as 4-probe, transmission line, Hall-effect measurements, etc. A carrier concentration change $\Delta N_c = 2.12\times10^{20}$ cm$^{-3}$ estimated from the gated measurements (see results and discussion section) places the operation of these devices between carrier levels of $(2.07 - 4.19)\times10^{20}$ cm$^{-3}$, which is still in the $n$-dominant operation region, however intentionally away from the high-loss ENZ (6-7×$10^{20}$ cm$^{-3}$) state, yet sufficiently near to capture a slow-light effect [5]. Modulated carrier concentration profile is implemented using the modified Thomas-Fermi approximation (MTFA) method characterized by the MTFA-associated screening length, $\lambda_{MTFA} \sim 3$ nm [49]. MTFA allows for the quantum-mechanical influence of an infinite potential barrier at the surface and has shown compliance for semiconductors with surface band bending at the nanoscale [49].

In an electrical capacitive stack with ITO being one of the electrodes, application of a drive voltage can place the capacitor into states of accumulation or depletion (inversion in ITO has not yet been demonstrated), thus changing the carrier concentration and hence the optical complex index. The optical property of the active material, here ITO, therefore changes significantly, resulting in strong optical modulation effects. In praxis, a $1/e$ decay length of about 5 nm has been measured before [50], and modulation effects have been experimentally verified over $1/e^2$ (~10 nm) thick films from the interface of the oxide and ITO [1]. In order to extract relevant parameters including the effective indices (real and imaginary parts, $n_{eff}$ and $\kappa_{eff}$) and confinement factors, $\Gamma$, we perform FEM eigenmode analysis for our structure (Fig. 3a, inset). The first order transverse magnetic (TM)-like mode is selected following the TM-optimized grating couplers in the fabricated device and the mode profiles indicate an increase in the light confinement with modulation by almost 2.03× which is aligned with results from our previous work as we operate sufficiently away from the ENZ point in the $n$-dominant region (Fig. 3a) [5,43,44].

Passive MZI designs are swept using waveguide lengths for both the arms of the interferometer structure and the output waveguide separately to find the passive optical losses in the taped out structure and determine the length dependent and independent losses. In order to ascertain insertion losses arising from the fabricated active device, we choose to carry out a transmission line measurement of the passive MZI structures varying lengths of the waveguides. Employing a cutback method, we can differentiate the length dependent and independent losses in the passive structure (Fig. 3b). The length independent losses (i.e. arising from the grating couplers, waveguide bends, Y-junctions, etc.) can aid the insertion loss calculations for the active device. We choose to vary the arm waveguide lengths of the MZI maintaining a symmetrical design and also independently varied the output waveguide length in the structure. Both the sweeps provide nearly the identical results in the static losses. We find the length independent static losses are about 22 dB for our MZI design accounting for all waveguide bends, grating couplers and Y-junctions. The Y-junctions used fair an average insertion loss of 0.28±0.02 dB each [51]. Previously we found similar TM grating coupler losses to be 8 dB/coupler on the same SOI platform [52]. Neglecting the fabrication variation that might arise from separate multi project runs, we can roughly estimate the waveguide bending losses as 0.6 dB/bend from our structure with all bending radii of 60 μm maintained constant throughout the design (i.e. to and from the Y-junctions for both MZI arms and after the output Y-junction).

Active phase shifter devices are also swept in length from sub-λ (1.4 μm) to λ-scale (3.5 μm) range for determining propagation and coupling losses for the active capacitive stack formed by the ITO-oxide-metal layers from the waveguide and passivation cladding layer interface (Fig. 3c). Cutback measurements reveal 1.6 dB/μm propagation loss in the active capacitive stack and an additional 1.3 dB/coupler loss from in/out coupling of the mode from the Si waveguide, while the passive contact for loss balancing (Fig. 1a, $L_b$) exhibits a 1.2 dB/μm propagation loss and 1.1 dB/coupler loss correspondingly. The corresponding overall insertion losses (IL) for the differently scaled modulators range from 6.7 – 10.1 dB. Note that the high loss per unit length in plasmonics is alleviated by an enhanced light–matter interaction enabling λ-short device lengths; thus the total IL is comparable to Si photonic MZIs.

## Results and Discussion

Our sweep of the active phase-shifter device length ($L_{ps}$) ranges from sub-λ (1.4 μm) to λ-scale devices (3.5 μm) [Fig. 1b]. Broadband spectral response is measured in C-band region (~30 nm, Fig. 4a); which is expected since the plasmonic resonance of the mode has a FWHM of 100's of nm. The spectral response of the device is determined by ITO dispersion and proximity to ENZ. For ultra–broadband applications (e.g. 100+ nm) ITO modulators for different spectral regions (e.g. $\Delta\lambda = 50$ nm) can be processed using different conditions as demonstrated earlier [6]. Functional capacitor traits in the measured bias range is observed from I-V measurements (Fig. 4b). The I-V measurements do not portray any breakdown of the gate oxide or saturation behavior of the MOS capacitor in the applied voltage range. Note, the gate oxide breakdown occurs at nearly 12 V which was confirmed in several devices on the same chip.

DC electro-optic transmission power tests and squared cosine fit (as dictated by MZI operating principle, see methods) result in an ER of ~3 dB to >8 dB, respectively (Fig. 4c). The measured $V_\pi L$ is just 95±2 V·μm and found to be rather constant across all device scaling. The quality of the fit symbolized by the coefficient of determination ($R^2$) is above 0.99 in all three cases. While these prototype devices require a relatively high voltage ($V_{\pi, \text{as-is}}$ ~26 V, $\epsilon_{r, Al_2O_3} = 11$, $t_{ox}$ = 20 nm), the $V_{\pi, \text{future-option}} < 2$ V can be reached using 5 nm high-k dielectrics (e.g. $Hf_3N_4$ [53]) in a push-pull configuration and trivial capacitor reductions during lithography. The results indicate a modal index change $\Delta n_{eff}$ of ~ 0.16 due to carrier modulation across the applied voltage range (Fig. 4d) and FEM eigenmode analysis (inset, Fig. 3a) reveals an index change in the active ITO material of about 0.6 (Fig. 4e) reflecting a ~2.03× increase in the modal confinement factor ($\Gamma$) corresponding to active biasing, slightly lower than previous modulators in ITO [3], and intentionally enabling lower insertion loss (IL). The effective index change with modulation can be expressed by a linear approximation with applied voltage as $\partial n_{eff}/\partial V_d \sim 7.98\times10^{-3}$ V$^{-1}$ ~ 1%V$^{-1}$. The material index change shows meager nonlinearity for negative biasing as we operate in the ITO material's $n$-dominant region sufficiently close to ENZ beneficially extracting proximal ENZ effects while staying away from the high-loss ENZ point to minimize IL concerns [40]. Note, the change in both the indices (material and effective) resemble a decrease in the corresponding indices as modulation assimilates to blue-shifts in device resonance, however, is barely resolvable in our single pass MZ configuration (Fig. 4a) but is well-known from any cavity based operations. Modal dispersion, $\partial n_{eff}/\partial \lambda$ calculated from eigenmode analysis indicates $1.12\times10^{-3}$ nm$^{-1}$ resonance shift in our MOS structure. The calculated group index in the modal structure is 4.1 indicating slow-light effects attribute to about 0.76 nm shift in the corresponding spectrum across applied bias of 20 V (Fig. 4a). However, our non-resonant MZI measurements do naturally not resolve this resonance shift given the plasmonic mode interferes in ascertaining the central resonant wavelength; this reported value is taken from the transmission spectra which can relinquish to various noise in the system (Fig. 4a).

To test the temporal response of these sub-λ-compact ITO MZI modulators, the frequency response ($S_{21}$) is obtained by generating a low power modulating signal (0 dBm) with a 50 GHz network analyzer; a bias-tee mixes DC voltage bias (6 V) with the RF signal (Fig. 5a). The RF output from the modulator is amplified using a broadband EDFA (~35 dB), an optical tunable filter enhances the signal integrity and reduces undesired noise by 20 dB. The modulated signal is collected by a high-speed photodetector ($\tau_{response}$ = 10's ps) with a single mode fiber. The -3 dB roll-off (small signal) shows a speed of 1.1 GHz (Fig. 5b), which matches estimations for the RC-delay given a capacitance of 213 fF (area = 42 μm$^2$) and a total resistance of 680 Ω, aligned with our earlier results [3]. Plasmonics enables reduced series resistance in device configurations as one of the contacts realized by metal is 'free' leading to small RC-delays [54], as shown here. A proportionate dynamic switching energy of ~2.1 pJ/bit characterizes the spectral broadband tradeoff in the sub-λ device. Comparative performance metrics across all device length scaling are shown where the goal is to minimize the value in all axes (Fig. 5c). The sub-λ device outperforms both longer counterparts in all metrics mainly benefitting from the reduced resistance and capacitance. Future improvements such as optimized contact placement, pad-overlay optimization, annealing, plasma treatment, deployment of high-k gate dielectrics including $t_{ox}$ scaling, and utilizing push-pull schemes can enable 10's of GHz operation requiring only a few fJ/bit. However, aJ/bit energy levels are likely not feasible in non-resonator schemes due to the tradeoff in spectral bandwidth (i.e. no cavity feedback) [40].

Indeed, novel electro-optic modulators allow in addition to signal amplitude modulation [57], but further allow designing efficient switches that can be assembled into cross-bard routers [58]. Indeed, capabilities of nanomanufacturing and hence the control at a nanoscale over traditional but also emerging materials [59,60] allows for a number of electro-optic functional performances, such as atomistic small switches, thus pushing for the ultimate small formfactors of modulators, such as demonstrated by the Leuthold group [61].

## Conclusion

In conclusion, here, we have experimentally demonstrated a spectrally broadband, extremely compact (sub-wavelength short), GHz-fast ITO-based Mach Zehnder Interferometer-based modulator in silicon photonics exhibiting a low V$_\pi$L of 95 V·μm enabled by a) efficient material modulation in ITO optimized for real-part index operation, b) a plasmonic hybrid mode enhancing the light-matter interaction, c) relatively low electrical resistance, and d) operating away from an optical resonance. This demonstration bears relevance, since ITO is a foundry-compatible material with a reduced barrier for co-integration. Unlike the crystal orientation sensitive LiNbO$_3$, ITO optoelectronics is synergistic to enhancing electrostatics known from transistor technology. Such micrometer-compact next generation and foundry-near phase-shifters and electro-optic modulators monolithically integrated with photonic integrated circuits, here silicon as an example, enable high PIC-component density valuable for applications such as data communication, signal processing, GHz-fast optical phased arrays for beam-steering applications, and neuromorphic photonics and optical information processing.

## Materials and Methods

### MZI Transfer Matrix

Different configurations in MZI structures can be studied using the transfer matrix model. Each sub-component of the MZI is represented by a vector or a matrix and the final outputs are obtained by multiplying the individual matrices. The matrix formulation of the electric field outputs of the Y-junction is:

$$\begin{bmatrix} E_{Y,Out1} \\ E_{Y,Out2} \end{bmatrix} = \begin{bmatrix} \sqrt{\delta_1} \\ \sqrt{\delta_2} \end{bmatrix} [E_{In}] \qquad (1)$$

where, $E_{In}$, $E_{Y,Out1}$, and $E_{Y,Out2}$ are the electric field phasors of the input, output of arm 1, and output of arm 2, respectively. The $\delta_1$ and $\delta_2$ are the output power ratio of arm 1 and arm 2 of the Y-junction relative to the input. For a lossless Y-junction with identical splitting, $\delta_1 = \delta_2 = \frac{1}{2}$. The outputs of the 50:50 directional coupler, ignoring loss, in matrix form is:

$$\begin{bmatrix} E_{DC,Out1} \\ E_{DC,Out2} \end{bmatrix} = \begin{bmatrix} \sqrt{1-\delta} & i\sqrt{\delta} \\ i\sqrt{\delta} & \sqrt{1-\delta} \end{bmatrix} \begin{bmatrix} E_{In1} \\ E_{In2} \end{bmatrix} \qquad (2)$$

where, $\delta$ is the power splitting ratio between the two arms. The imaginary unit $i$ represents the π/2 phase shift between the direct and cross coupled inputs. The propagation in the arms of the interferometer is modeled by modifying the amplitude and the phase of the phasors.

$$\begin{bmatrix} E_{Out,arm1} \\ E_{Out,arm2} \end{bmatrix} = \begin{bmatrix} \exp(-i\phi_1 - \alpha_1 L_1/2) & 0 \\ 0 & \exp(-i\phi_2 - \alpha_2 L_2/2) \end{bmatrix} \begin{bmatrix} E_{In,arm1} \\ E_{In,arm2} \end{bmatrix} \qquad (3)$$

where, $\phi_1$ and $\phi_2$ is the total phase shift in arm 1 and arm 2, $\alpha_1$ and $\alpha_2$ is the optical propagation loss in the respective arms, and $L_1$ and $L_2$ are the total lengths of each arm. The terms $E_{In,arm1}$ and $E_{In,arm2}$ are the input fields at the two arms of the interferometer, respectively, and $E_{Out,arm1}$ and $E_{Out,arm2}$ are the fields after the two phase shifting regions, respectively.

Phase shift differences between the two paths of the interferometer, in silicon MZIs, can be introduced by an imbalance in length, thermal tuning, and modulation by the plasma dispersion effect. The imbalance can be unintentional, for instance due to variations in fabrication or intentional to create a measurable free spectral range and facilitate phase shift measurements. The total phase shift and loss can be divided into these three individual components, thus, the total phase shift for arm 1 and arm 2 can be expressed as:

$$\phi_1 = \frac{2\pi}{\lambda} \left[ n_{eff} L_{nm,1} + n_{eff}(V) L_{active,1} + n_{eff}(T) L_{thermal,1} \right] \qquad (5)$$

$$\phi_2 = \frac{2\pi}{\lambda} \left[ n_{eff} L_{nm,2} + n_{eff}(V) L_{active,2} + n_{eff}(T) L_{thermal,2} \right] \qquad (6)$$

Where, $L_{nm,1}$ and $L_{nm,2}$ are the non-modulated lengths of each arm, $L_{active,1}$ and $L_{active,2}$ are the electrically modulated lengths of each arm, and $L_{thermal,1}$ and $L_{thermal,2}$ are the thermally modulated lengths of each arm. The effective index of refraction is a function of voltage for electrical modulation and a function of temperature for thermal modulation. With this partition, the total length of an arm $L$ is the sum of the lengths $L_{nm}$, $L_{active}$, and $L_{thermal}$ of that arm. Finally, the total transmission can be obtained by multiplying the matrices together. For a 1×1 MZI, the output electric field is obtained as:

$$[E_{Out}] = [Y]_{out} [MZ] [Y]_{in} E_{in} \qquad (7)$$

Where, [Y] represents the vector of the Y-junctions at the output and input, and [MZ] is the 2×2 matrix for the phase shift in the arms of the interferometer. The output intensity $I_{out}$ is then obtained by the square of the absolute value of the output electric field. The optical intensity in decibels is:

$$I_{out,dB} = 10 \log_{10}(|E_{Out}|^2) \tag{8}$$

A simplified expression for the electric field at the output of the 1×1 MZI can be obtained by considering only one type of phase changing mechanism and assuming equal physical path lengths. The electric field output for this special case involves both an amplitude modulation and a phase modulation component. For intensity modulation, the amplitude modulation is desired but the phase modulation is unwanted. On the other hand, for phase shift keying modulation formats, the amplitude modulation that occurs is undesired.

$$E_{Out} = E_{in} \cos\left(\frac{\Delta \beta L}{2}\right) exp\left[i\left(\frac{\Delta \beta L}{2}\right)\right] \tag{9}$$

The argument of the cosine term corresponds to the amplitude modulation as a byproduct of the phase variation whereas the exponential term corresponds to the actual change in the phase of the propagating wave. Similarly, a simplified expression for the intensity at the output of the 1×1 MZI can be obtained by employing the exponential-hyperbolic and Euler's relation. The transmission of the MZI is:

$$T = \frac{1}{2}\left[1 + sech\left(\frac{\Delta \alpha L}{2}\right) \cos\left(\frac{2\pi}{\lambda} \Delta n_{eff} L\right)\right] \tag{10}$$

where, $\Delta \alpha$ is the difference in loss between the two arms. The argument in the cosine term is the relative phase shift leading to the cosine term evaluating to 1 for constructive interference, and -1 for destructive interference.

**ITO ion beam deposition**

The 10 nm film of ITO was deposited on top of one of the passive waveguides (one arm of the MZ) on SOI substrate at room temperature via ion beam deposition (IBD) using the 4Wave IBD/BTD cluster sputter deposition system following a passivation oxide ALD layer of 5 nm. An RF ion gun focused Ar ions onto substrate targets of ITO. The ITO target stoichiometry is 90 wt % $In_2O_3$ / 10 wt % $SnO_2$. A small flow of $O_2$ (2 sccm) was used. The process used an Ar flow rate of 20 sccm, beam voltage of 600 V, beam current of 220 mA and acceleration voltage of 150 V. The sample was set at an angle of 115° and rotated at 10 rpm to ensure smooth profile. The deposition rate used was 0.77 Å/s. The temperature for the process was 20°C to refrain annealing effects. The base vacuum used was $2\times10^{-8}$ Torr and the deposition uniformity was confirmed as 1.5% (1σ) over a 190 mm diameter.

**Fabrication processes**

The pattern transfers were done in e-beam lithography (EBL) using the Raith VOYAGER tool with PMMA based photoresists and MIBK1:3IPA developer for 60 sec. 47 nm of Au for contacts was deposited using an e-beam evaporation system (CHA Criterion) as it has reasonably low ohmic loss at near IR wavelengths. An additional 3 nm adhesion layer of Ti was used in the contacts. The $Al_2O_3$ oxide was deposited using the atomic layer deposition (ALD) technique as it provides reliable performance characteristics. The Fiji G2 ALD tool was used at low temperature settings (100°C) for 100 cycles to deposit about 20 nm of $Al_2O_3$ to ensure higher film quality and to avoid any annealing effects to the ITO. Filmetrics F20-UV system was used to characterize $Al_2O_3$ deposition rate. As the ALD tool does not allow patterned processing due to chamber contamination concerns, an etch step was required on top of the Si contact to remove the oxide over it for electrical probing. We used a rather slow wet etch process for $Al_2O_3$ using an MF319 solution in the contact pad area (See supplementary information). MF319 contains tetramethylammonium hydroxide (TMAH) which reacts with the Al and can etch the oxide thereof.

**Spectroscopic ellipsometry**

The J. A. Woollam M-2000 DI spectroscopic ellipsometer was used to characterize the optical constants of the deposited ITO thin films as it can provide fast and accurate thin film characterization over a wide spectroscopic range. Variable angle spectroscopic elliposometric (VASE) data was collected from three different angles of 65°, 70° and 75°. First, the transparent region was fitted using Cauchy model to find the closest thickness value for ITO thin film with the deposited value and was kept fixed. Then, the data was fitted using B-spline model and subsequently the fitting region was expanded from transparent region to the entire wavelength region. Afterwards the data was re-parameterized using different oscillators in GenOsc, i.e. Drude, Lorentz and Tauc-Lorentz oscillators to find relevant optical constants (See supplementary information). The Tauc-Lorentz model has been reported to provide excellent fitting to various TCO materials [55,56]. A mean square error (MSE) value of 2.001 and uniqueness of the fitted thickness parameter coupled with matching the thickness from deposition reinforces the fit.

## Additional Information

**Supplementary Information**

Refer to accompanying supplementary information.

**Competing interests**

The authors declare no competing interests.

# Figures and Tables

**Table 1:** Figure of merit (FOM) comparison for recent Mach Zehnder interferometric schemes with different active materials and waveguide structures

| Structure/Material | $V_\pi L$ (V.µm) | Ref. |
|---|---|---|
| Si Wrapped around-pn | 140,000 | [14] |
| Ferroelectric domain inverted coplanar waveguide LiNbO$_3$ | 120,000 | [15] |
| Si Wrapped around-pn | 110,000 | [14] |
| Domain inverted push-pull LiNbO$_3$ | 90,000 | [16] |
| Dual driven coplanar waveguide LiNbO$_3$ | 80,000 | [17] |
| Si Vertical-pn | 40,000 | [18] |
| Bulk LiNbO$_3$ physical limit | 36,000 | [19] |
| Si pipin | 35,000 | [20] |
| Si Lateral-pn | 28,000 | [21] |
| Si Lateral-pn | 27,000 | [22] |
| Si pn-depletion | 24,000 | [23] |
| Doping optimized Si | 20,500 | [24] |
| Si Self-aligned-pn | 18,600 | [25] |
| Integrated thin film LiNbO$_3$ on insulator | 18,000 | [26] |
| Si pin | 13,000 | [27] |
| Silicon-organic hybrid (SOH) | 9,000 | [28] |
| Si Lateral-pn | 8,500 | [29] |
| Si Projection MOS | 5,000 | [30] |
| III-V Multiple Quantum Wells (MQW) | 4,600 | [31] |
| SOH | 3,800 | [32] |
| GaAs/AlGaAs | 2,100 | [33] |
| Hybrid Si MQW | 2,000 | [34] |
| InGaAlAs/InAlAs MQW | 600 | [35] |
| ITO MOS | 520 | [3] |
| Si p$^+$-i-n$^+$ | 360 | [36] |
| ITO MOS Plasmonic | 95 | This work |
| EO Polymer Plasmonic | 70 | [37] |
| ITO Lateral MOS | 63 | [7] |
| Liquid crystals with SOH slot/ all-plasmonic polymer | 60 | [38], [39] |

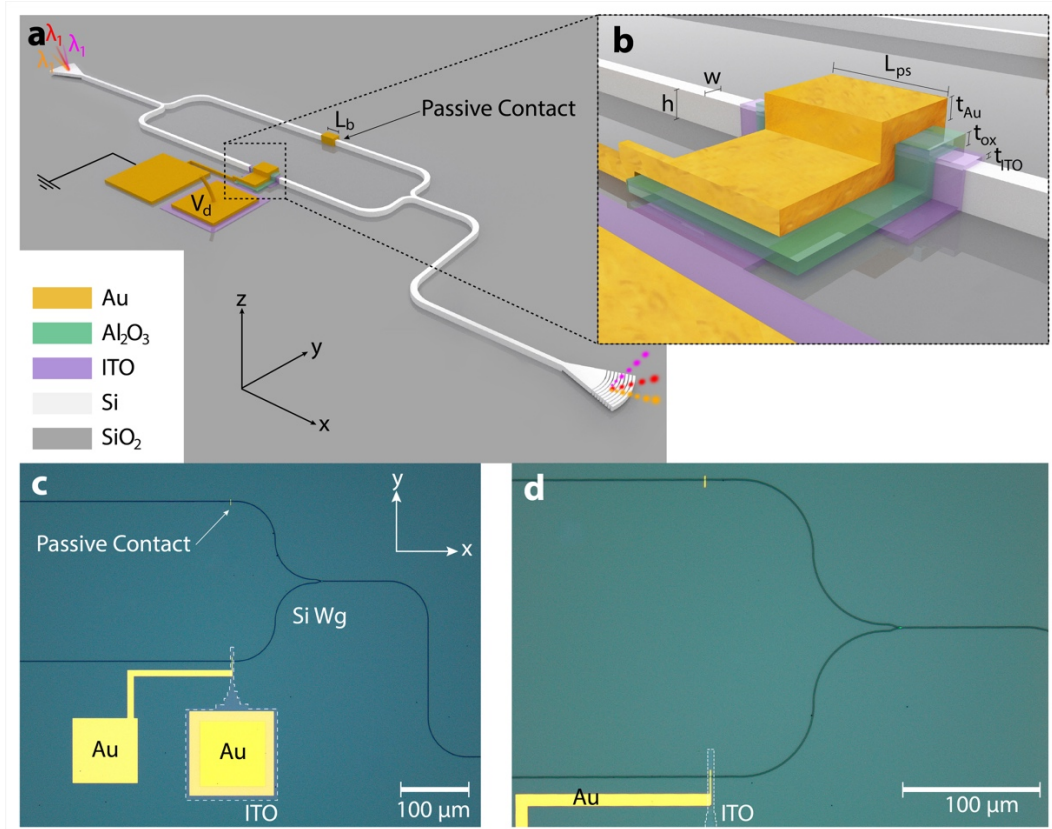

**Figure 1. ITO-based plasmonic Mach-Zehnder interferometric modulator on Si photonic platform. a.** Perspective view of the Mach-Zehnder structure with the active biasing contacts showcasing the broadband operation enabled by simultaneous operation at different wavelengths, $\lambda_1$, $\lambda_2$, $\lambda_3$, etc. Passive metallic contact, $L_b$ is for the field loss balancing. **b.** The active modulation region is highlighted. Relevant parameters include: Phase shifter length, $L_{ps}$ is swept from sub-$\lambda$ to $\lambda$-scale lengths; thickness of the deposited Au, gate oxide, $Al_2O_3$, and ITO thin films: $t_{Au}$ = 50 nm, $t_{ox}$ = 20 nm, $t_{ITO}$ = 10 nm. The Si waveguide dimensions are: $w$ = 500 nm, and $h$ = 220 nm. Image not drawn to scale. **c.** Optical microscope image of the fabricated device; the patterned ITO thin film region is highlighted with white dashes. **d.** Zoomed in optical micrograph of the sub-wavelength device section formed by patterned ITO (white dashed) and top metallic (Au) contact separated by an atomic layer deposited gate oxide film everywhere on the chip.

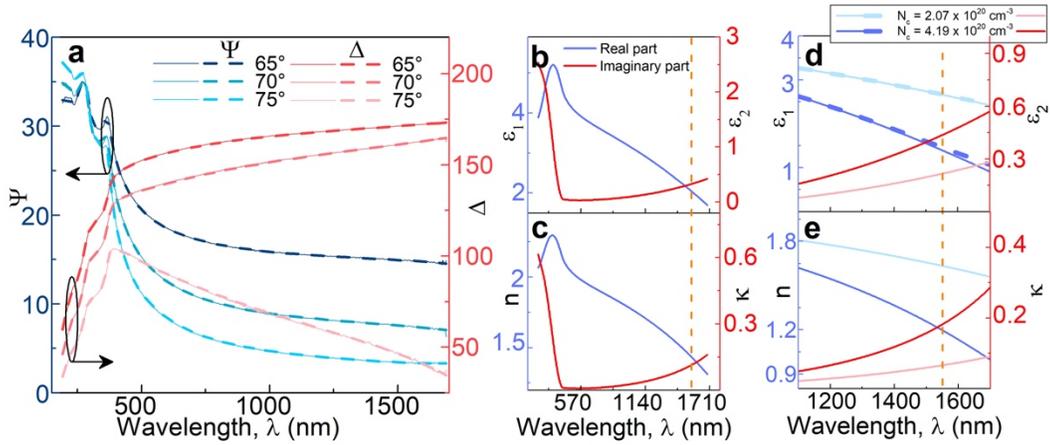

**Figure 2. Ion beam deposited (IBD) ITO material optical properties. a.** Variable angle spectroscopic ellipsometric data, Ψ and Δ with respect to wavelength, λ for different incident angles of 65º, 70º and 75º. Actual measurement data represented by solid lines and relevant fitted model shown with dashes. **b.** Corresponding real and imaginary parts of the permittivity, $\varepsilon_1$ and $\varepsilon_2$; and **c.** Real and imaginary parts of the complex refractive index, $n$ and $\kappa$ vs. wavelength, λ from spectroscopic ellipsometry. **d.** Real and imaginary parts of the permittivity, $\varepsilon_1$ and $\varepsilon_2$; and **(e)** Real and imaginary parts of the index, $n$ and $\kappa$ dispersion in the NIR region obtained from Drude formulation using ellipsometric fit results as a function of the carrier concentration, $N_c$ variation arising from active gating in modulator operation. Carrier concentration levels, $N_c$ of $2.07 \times 10^{20}$ cm$^{-3}$ and $4.19 \times 10^{20}$ cm$^{-3}$ are found from active modulation results and correspond to the modulator ON and OFF states, respectively. Intended operating wavelength in the telecom C-band (λ = 1550 nm) is highlighted with orange dashed line.

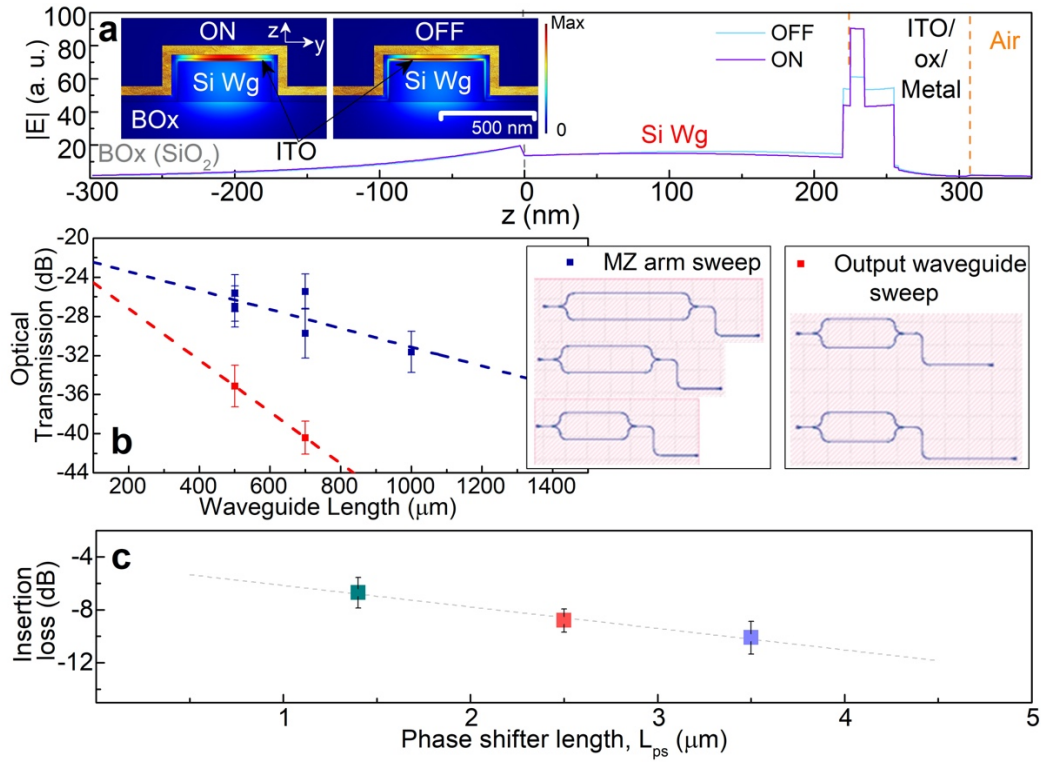

**Figure 3. Modal analysis and passive (un-modulated) propagation results. a.** Electric field distribution in the cross-sectional structure for a z-cutline along the central region of the Si waveguide (width) corresponding to the ON and OFF states of operation. The respective mode profiles are shown as inset. **b.** Optical transmission (dB) through the passive MZI structures for waveguide length sweeps. Both the arms of the MZI and the output waveguide are swept in length independently for transmission line measurements to characterize the length dependent and independent optical losses. **c.** Insertion losses of the fabricated active devices against their respective lengths formulating cutback analysis.

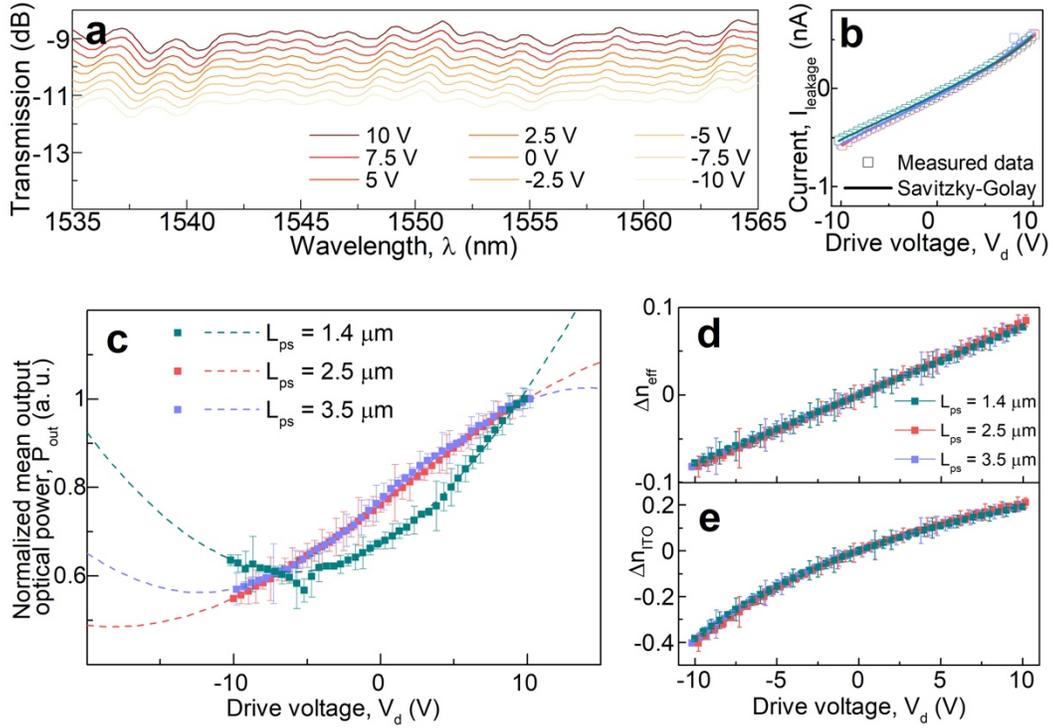

**Figure 4. DC electro-optic modulation results. a.** Gated transmission characteristics of the fabricated Mach-Zehnder modulator ($L_{ps}$ = 2.5 μm) exhibiting broadband performance in the telecom C band. **b.** I-V measurements of the fabricated devices. A Savitzky-Golay smoothing function has been applied on the measured data to showcase the I-V characteristics of the device. **c.** Normalized optical output power (a. u.) vs. drive voltage, $V_d$ (V) corresponding to different phase shifter lengths, $L_{ps}$. Squared cosine ($\cos^2$(arg)) fits dictated by Mach-Zehnder operating principle are shown with dashed lines. **d.** Induced modal effective index variation, $\Delta n_{eff}$; and **e.** Induced ITO material index change, $\Delta n_{ITO}$; under MOS-capacitive gating arising from applied drive voltage, $V_d$ (V) across different phase shifter length ($L_{ps}$) scaling.

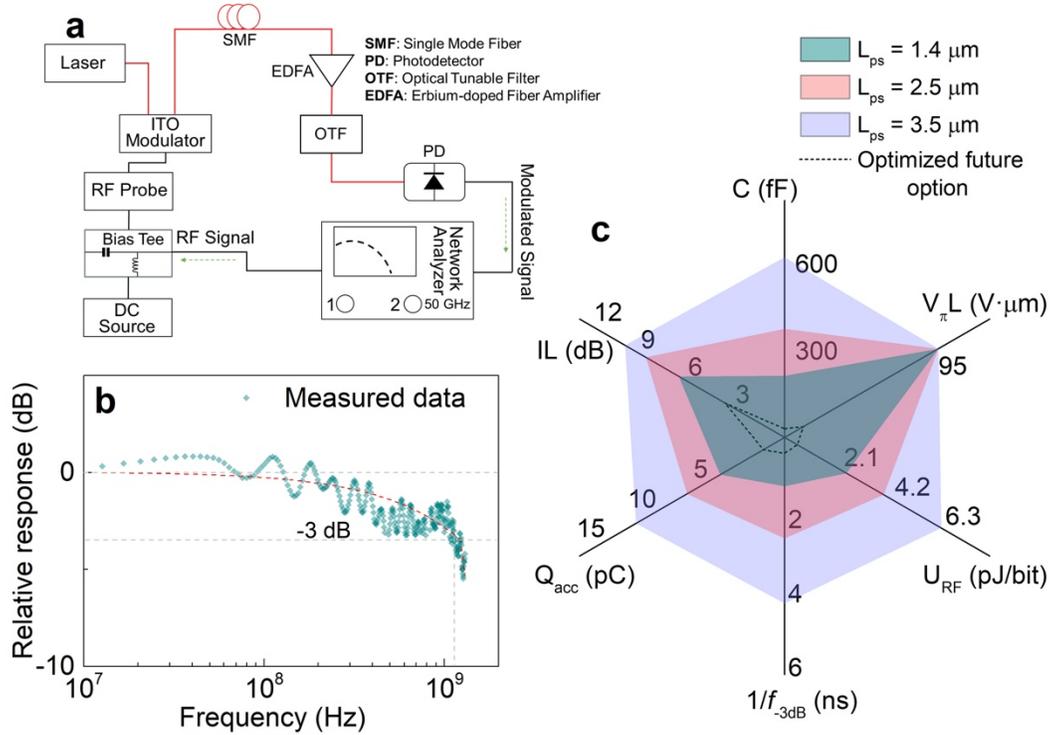

**Figure 5. RF electro-optic modulation measurements and performance metrics across different phase-shifter scaling. a.** Measurement setup for the electro-optic frequency response showing electrical connections (black) and optical paths (red) using single mode fibers. **b.** Relative frequency response, $S_{21}$ (dB) exhibiting a -3 dB cutoff speed at around 1.1 GHz. **c.** Different performance metrics including capacitance, C (fF); figure of merit, $V_\pi L$ (V·μm); dynamic switching energy; $U_{RF}$ (pJ/bit); -3dB cutoff latency, $1/f_{-3dB}$ (GHz); accumulated charge, $Q_{acc}$ (pC); and insertion losses, IL (dB) for all 3 different device length scales. Performance metrics for an optimized future option of the current ITO paradigm enabling CMOS-low drive voltages are shown with a dashed line. A smaller value on each axis denotes a higher performance with a modulator performance merit indicated by a smaller area.

17